\documentclass[twocolumn,twocolappendix]{aastex631}
\usepackage{CJKutf8}
\usepackage{amsmath}

\pdfoutput=1

\begin{document}
\begin{CJK}{UTF8}{gbsn}
\title{The magnetic field in quiescent star-forming filament G16.96+0.27}

\author[0000-0002-2826-1902]{Qi-Lao Gu (顾琦烙)}\thanks{E-mail: qlgu@shao.ac.cn}
\affiliation{Shanghai Astronomical Observatory, Chinese Academy of Sciences \\
No.80 Nandan Road, Xuhui, Shanghai 200030, People's Republic of China}

\author[0000-0002-5286-2564]{Tie Liu (刘铁)}\thanks{E-mail: liutie@shao.ac.cn}
\affiliation{Shanghai Astronomical Observatory, Chinese Academy of Sciences \\
No.80 Nandan Road, Xuhui, Shanghai 200030, People's Republic of China}

\author[0000-0003-3540-8746]{Zhi-Qiang Shen (沈志强)}\thanks{E-mail: zshen@shao.ac.cn}
\affiliation{Shanghai Astronomical Observatory, Chinese Academy of Sciences \\
No.80 Nandan Road, Xuhui, Shanghai 200030, People's Republic of China}

\author{sihan Jiao (焦斯汗)}
\affiliation{National Astronomical Observatories, Chinese Academy of Sciences \\
A20 Datun Road, Chaoyang, Beijing 100101, People's Republic of China}

\author[0000-0001-7032-632X]{Julien Montillaud}
\affiliation{Université de Franche-Comté, CNRS, Institut UTINAM, OSU THETA, F-25000 Besan\c{c}on, France}

\author[0000-0002-5809-4834]{Mika Juvela}
\affiliation{Department of Physics, PO box 64, FI-00014, University of Helsinki, Finland}

\author[0000-0003-2619-9305]{Xing Lu (吕行)}
\affiliation{Shanghai Astronomical Observatory, Chinese Academy of Sciences \\
No.80 Nandan Road, Xuhui, Shanghai 200030, People's Republic of China}

\author[0000-0002-3179-6334]{Chang Won Lee}
\affiliation{Korea Astronomy and Space Science Institute, 776 Daedeokdae-ro, Yuseong-gu, Daejeon 34055, Republic of Korea}
\affiliation{University of Science and Technology, Korea, 217 Gajeong-ro, Yuseong-gu, Daejeon 34113, Republic of Korea}

\author[0000-0002-4774-2998]{Junhao Liu(刘峻豪)}
\affiliation{National Astronomical Observatory of Japan, 2-21-1 Osawa, Mitaka, Tokyo, 181-8588, Japan}

\author[0000-0001-8077-7095]{Pak Shing Li}
\affiliation{Shanghai Astronomical Observatory, Chinese Academy of Sciences \\
No.80 Nandan Road, Xuhui, Shanghai 200030, People's Republic of China}

\author[0000-0001-8315-4248]{Xunchuan Liu (刘训川)}
\affiliation{Shanghai Astronomical Observatory, Chinese Academy of Sciences \\
No.80 Nandan Road, Xuhui, Shanghai 200030, People's Republic of China}

\author[0000-0002-6773-459X]{Doug Johnstone}
\affiliation{NRC Herzberg Astronomy and Astrophysics, 5071 West Saanich Rd, Victoria, BC, V9E 2E7, Canada}
\affiliation{Department of Physics and Astronomy, University of Victoria, Victoria, BC, V8P 5C2, Canada}

\author[0000-0003-4022-4132]{Woojin Kwon}
\affiliation{Department of Earth Science Education, Seoul National University, 1 Gwanak-ro, Gwanak-gu, Seoul 08826, Republic of Korea}
\affiliation{SNU Astronomy Research Center, Seoul National University, 1 Gwanak-ro, Gwanak-gu, Seoul 08826, Republic of Korea}
\affiliation{The Center for Educational Research, Seoul National University, 1 Gwanak-ro, Gwanak-gu, Seoul 08826, Republic of Korea}

\author[0000-0003-2412-7092]{Kee-Tae Kim}
\affiliation{Korea Astronomy and Space Science Institute, 776 Daedeokdae-ro, Yuseong-gu, Daejeon 34055, Republic of Korea}
\affiliation{University of Science and Technology, Korea, 217 Gajeong-ro, Yuseong-gu, Daejeon 34113, Republic of Korea}

\author[0000-0002-8149-8546]{Ken'ichi Tatematsu}
\affiliation{Nobeyama Radio Observatory, National Astronomical Observatory of Japan,
National Institutes of Natural Sciences\\
Nobeyama, Minamimaki, Minamisaku, Nagano 384-1305, Japan}
\affiliation{Astronomical Science Program,
Graduate Institute for Advanced Studies, SOKENDAI\\
2-21-1 Osawa, Mitaka, Tokyo 181-8588, Japan}

\author[0000-0002-7125-7685]{Patricio Sanhueza}
\affiliation{National Astronomical Observatory of Japan, 2-21-1 Osawa, Mitaka, Tokyo, 181-8588, Japan}
\affiliation{Astronomical Science Program,
Graduate Institute for Advanced Studies, SOKENDAI\\
2-21-1 Osawa, Mitaka, Tokyo 181-8588, Japan}

\author{Isabelle Ristorcelli}
\affiliation{IRAP, Universit\'{e} de Toulouse, CNRS, 9 avenue du Colonel Roche, BP 44346, 31028 Toulouse Cedex 4, France}

\author[0000-0003-2777-5861]{Patrick Koch}
\affiliation{Academia Sinica Institute of Astronomy and Astrophysics, No. 1, Section 4, Roosevelt Road, Taipei 10617, Taiwan (R.O.C.)}

\author[0000-0003-2384-6589]{Qizhou Zhang}
\affiliation{Center for Astrophysics | Harvard \& Smithsonian\\
60 Garden Street, Cambridge, MA 02138, USA}

\author[0000-0002-8557-3582]{Kate Pattle}
\affiliation{Department of Physics and Astronomy, University College London, Gower Street, London WC1E 6BT, United Kingdom}

\author[0000-0001-9304-7884]{Naomi Hirano}
\affiliation{Academia Sinica Institute of Astronomy and Astrophysics, No. 1, Section 4, Roosevelt Road, Taipei 10617, Taiwan (R.O.C.)}

\author[0000-0001-5403-356X]{Dana Alina}
\affiliation{Department of Physics, School of Science and Technology, Nazarbayev University, Astana 010000, Kazakhstan}
\affiliation{IRAP, Université de Toulouse CNRS, UPS, CNES, F-31400 Toulouse, France}

\author[0000-0002-9289-2450]{James Di Francesco}
\affiliation{Department of Physics and Astronomy, University of Victoria, Victoria, BC, V8W 2Y2, Canada}
\affiliation{2 NRC Herzberg Astronomy and Astrophysics, 5071 West Saanich Road, Victoria, BC, V9E 2E7, Canada}


\begin{abstract}
We present 850 $\rm\mu m$ thermal dust polarization observations with a resolution of 14.4\arcsec ($\sim0.13$ pc) towards an infrared dark cloud G16.96+0.27 using JCMT/POL-2. The average magnetic field orientation, which roughly agrees with the larger-scale magnetic field orientation traced by the \textit{Planck} 353 GHz data, is approximately perpendicular to the filament structure. The estimated plane-of-sky magnetic field strength is $\sim96$~$\rm\mu G$ and $\sim60$~$\rm\mu G$ using two variants of the Davis-Chandrasekhar-Fermi methods. We calculate the virial and magnetic critical parameters to evaluate the relative importance of gravity, the magnetic field, and turbulence. The magnetic field and turbulence are both weaker than gravity, but magnetic fields and turbulence together are equal to gravity, suggesting that G16.96+0.27 is in a quasi-equilibrium state. The cloud-magnetic-field alignment is found to have a trend moving away from perpendicularity in the dense regions, which may serve as a tracer of potential fragmentation in such quiescent filaments.
\end{abstract}



\section{Introduction} \label{sec:intro}
Filaments are ubiquitous in the interstellar medium \citep[e.g.,][]{Myers2009, Arzoumanian2011, Andre2010} with chains of dense cores embedded \citep[e.g.,][]{Zhang2009, Andre2014, Konyves2015, Tafalla&Hacar2015, Morii2023}, indicating that the filamentary structure might be an important stage in the star formation process \citep{LiuBb2012, Lu2018}. The details regarding how filaments fragment into dense prestellar cores and further evolve to form protostars are still under debate. Specifically, the role that the magnetic field plays during this process remains far from being fully understood \citep{Crutcher2012, LiHB2014, Pattle2023}.

\begin{figure*}[ht!]
\centering
\includegraphics[scale=1.3]{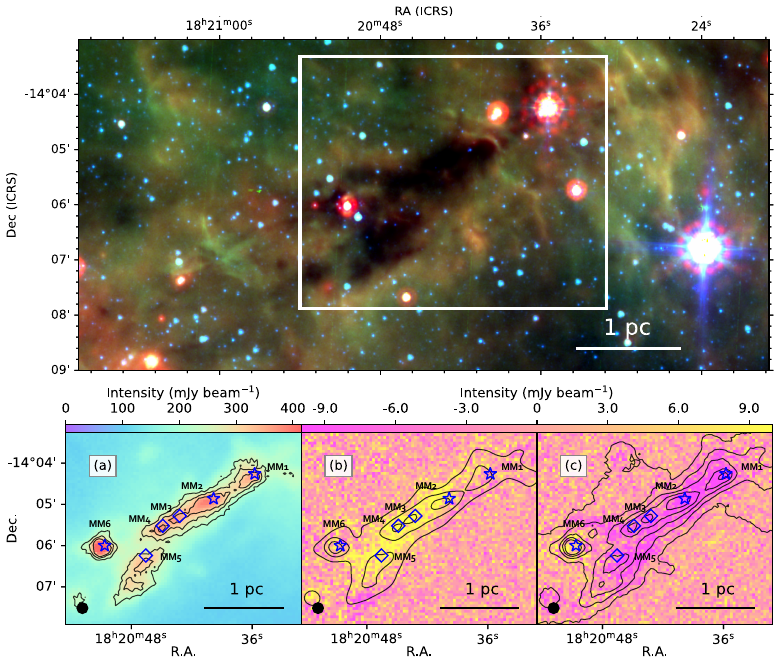}
\caption{\em{\textbf{Upper}: Spitzer infrared RGB map towards G16.96+0.27 (R: 24~$\mu$m; G: 8~$\mu$m; B: 5.8~$\mu$m). A 1-pc scale bar is shown in the lower right corner. The white box marks the region of lower panels \textbf{Lower (a-c)}: JCMT/POL-2 850~$\mu$m Stokes I, Q and U maps of G16.96+0.27, and Stokes I shown here is a combination of SCUBA-2 850~$\mu$m and deconvolved \textit{Planck} 353~GHz data using the J-comb algorithm \citep{Jiao2022} considering the large scale flux. Color bars are shown on the top of the lower panels. The details can be found in Appendix~\ref{sec:j-comb}. MM1 to MM6 mark the possible fragments observed at the resolution of 14.4$\arcmin$, star and diamond symbols represent protostellar and starless cores, respectively. Beams and the 1-pc scale bars are shown in the left and right corners, respectively. Contours in (a) show the intensity of 450~$\mu$m Stokes I at levels of [240, 480, 720]~mJy~beam$^{-1}$ with an average rms noise level of 44~mJy~beam$^{-1}$. The rms noise of the 450~$\mu$m Stokes \textit{Q} and \textit{U} maps is $\sim$41~mJy~beam$^{-1}$, which is not good enough for effective utilization, except in this figure, we do not show any other 450~$\mu$m results in this paper, and hereafter \textit{I}, \textit{Q} and \textit{U} refer to 850~$\mu$m data only. Contours in (b) show the column density generated from the \textit{Herschel} data by the J-comb algorithm \citep{Jiao2022} at levels of [2.22, 3.22, 4.22]$\times10^{22}$~cm$^{-2}$. Contours in (c) show the intensity of combined 850~$\mu$m Stokes I using J-comb algorithm at levels of [150, 200, 250, 300, 350]~mJy~beam$^{-1}$ with an average rms noise level of 5.3~mJy~beam$^{-1}$.}}
\label{fig:iqu}
\end{figure*}

Recent state-of-the-art ideal magnetohydrodynamic (MHD) simulations of large-scale filamentary cloud formation and evolution \citep[e.g.,][]{LiPS2019} suggest that a strong magnetic field perpendicular to the filament can support the filamentary structure and guide gas flow along the field onto the main cloud. Observationally, \cite{LiHB2015} found that the magnetic field orientation does not change much over $\sim100$ to $\sim0.01$ pc scale in the filamentary cloud NGC6334, suggesting self-similar fragmentation regulated by the magnetic field. Within nearby Gould Belt Clouds (with distances smaller than 500~pc), the parallel-to-perpendicular trend of cloud-field alignment (the offset between the magnetic field orientation and the molecular cloud long axis) with increasing density indicates that these clouds may have formed from the accumulation of material along the field lines \citep{PlanckXXXV}. With high-resolution submillimeter polarization observations, \cite{LiuT2018G35} found that in the massive infrared dark cloud (IRDC) G35.39-0.33 the magnetic field is roughly perpendicular to the densest part of the main filament but tends to be parallel with the gas structure in more diffuse regions. \cite{Soam2019} and \cite{Tang2019} found the magnetic field lines are more pinched by gravitational collapse at the core scale in the more evolved filamentary IRDC G34.42+0.24, where UC HII regions have formed. \cite{Ching2022} reported that a strong magnetic field shapes the main filament and subfilaments of the DR21 region. These results align with the simulations, suggesting that the magnetic field is dynamically important in the star formation process. However, active star formation has already occurred in these filamentary clouds, and feedback from star formation may have changed the initial magnetic field. Therefore observations of more quiescent clouds are required to investigate the role of the magnetic field in core formation inside filaments.

G16.96+0.27 is one of the brightest filaments in the JCMT SCOPE survey \citep{LiuT2018SCOPE} and is located at a distance of 1.87~kpc, embedded with few protostellar and starless cores \citep{Kim2020, Tatematsu2021, Mannfors2021}. As shown in the upper panel of Figure~\ref{fig:iqu}, G16.96+0.27 has a simple filamentary structure and is dark at the infrared wavelengths. So it has not been illuminated by protostars in the infrared band, suggesting it is a quiescent filament at the very early stage of the star formation process. This makes G16.96+0.27 an ideal target to study the magnetic field at the early stage of star formation. Here we use our 850 $\rm\mu m$ JCMT/POL-2 thermal dust polarization observations towards G16.96+0.27 to investigate the properties of the magnetic field inside a quiescent IRDC.

\begin{figure*}[ht!]
\centering
\includegraphics[scale=0.65]{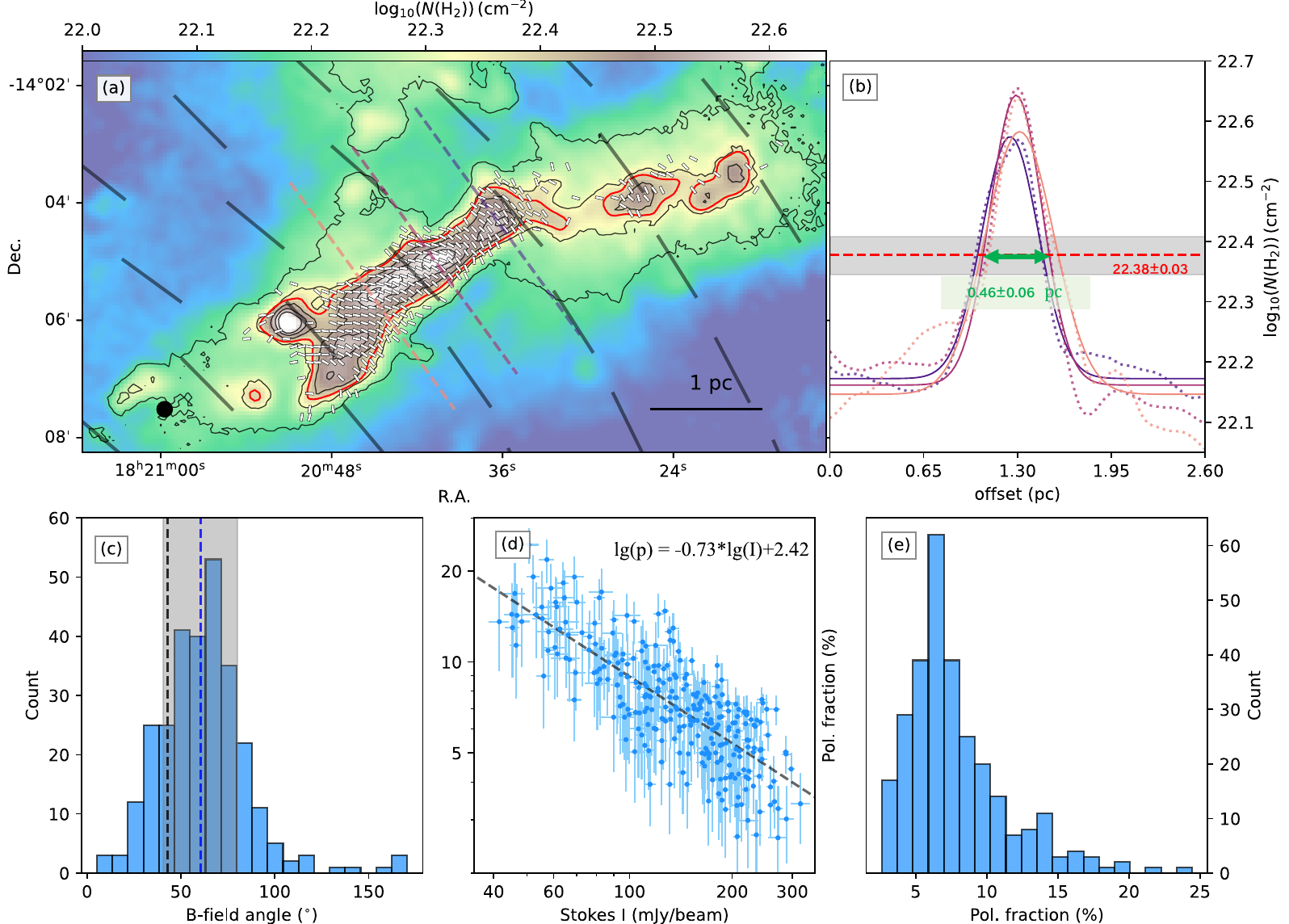}
\caption{\em{\textbf{(a)}: magnetic field orientations of G16.96+0.27 overlaid on the $N(\rm H_2)$ map with black contours showing 850~$\mu$m Stokes I levels of [150, 200, 250, 300, 350]~mJy~beam$^{-1}$. The short white and long black segments represent the magnetic field inferred from JCMT/POL-2 850~$\mu$m observation and \textit{Planck} 353~GHz data, respectively. The 14.4 $\arcsec$ beam size and a 1-pc scale bar are shown in the left and right lower corners. The dashed lines mark the cross-sections used for $N(\rm H_2)$ profile fitting shown in \textbf{(b)}, and the red contours represent the average FWHM $N(\rm H_2)$ value of $\sim2.39\times10^{22}\rm\,cm^{-2}$. \textbf{(b)}: Gaussian fittings of $N(\rm H_2)$ profiles from the three cross-sections in \textbf{(a)}, the offset is counted from northeast to southwest. Dashed and solid lines represent the observed data and best-fitting results, respectively. The red horizontal line marks the average FWHM $N(\rm H_2)$ and the grey region shows the uncertainty. The green double-headed arrow shows the average FWHM value of $0.46\pm0.06$~pc. \textbf{(c)}: Distribution of magnetic field orientations, the blue dashed line represents the average value and the grey region marks the standard deviation range. The black dashed line marks the average value of magnetic field orientations inferred from \textit{Planck} 353~GHz data. \textbf{(d)}: Polarization fraction vs. initial Stokes I (not the combined one using J-comb algorithm), the dashed line shows the power-law fit, and the best-fit parameters are shown in the top right corner. \textbf{(e)}: Distribution of polarization fraction.}}
\label{fig:pol_i}
\end{figure*}

This paper is organized as follows: in Section~\ref{sec2}, we present our JCMT/POL-2 850 $\rm\mu m$ observations; in Section~\ref{sec3}, we show the results from our observations and calculate the magnetic field strength; in Section~\ref{sec4}, we discuss the equilibrium state and multiscale cloud-field alignment; and we provide a summary in Section~\ref{sec5}.

\section{Observations} \label{sec2}
G16.96+0.27 was observed 19 times from 2020 August to 2020 October (project code: M20BP043; PI: Tie Liu) using SCUBA-2/POL-2 DAISY mapping mode \citep{Holland2013, Friberg2016, Friberg2018} under Band 2 weather conditions (0.05 $\textless\tau_{225}\textless$ 0.08, where $\tau_{225}$ is the atmospheric opacity at 225~GHz), with a total integration time of $\sim$12.8~hr. The effective beam size is $14.4\arcsec$ at 850~$\mu$m \citep{Mairs2021}, corresponding to $\sim$ 0.13~pc at a distance of 1.87~kpc. 

The raw data were reduced using the \textit{pol2map} routine of the STARLINK \citep{Currie2014} package, SMURF \citep{Chapin2013} with the 2019 August instrumental model\footnote{The details can be found in \url{https://www.eaobservatory.org/jcmt/2019/08/new-ip-models-for-pol2-data/}}, following the same procedures as described in \cite{Gu2024}. The final Stokes \textit{I}, \textit{Q} and \textit{U} maps are in units of pW with a pixel size of 4$\arcsec$, and are converted into the unit of Jy~beam$^{-1}$ by applying the 850~$\mu$m flux conversion factor (FCF) of 668~Jy~beam$^{-1}$ pW$^{-1}$ \citep{Mairs2021}. For the following analysis, we regrid these maps to a pixel size of 8$\arcsec$ for a balance of good S/N level and enough data points. The rms noise levels of background regions are $\sim$5.3 mJy~beam$^{-1}$ in the \textit{I} map. and $\sim$4.0~mJy~beam$^{-1}$ in the \textit{Q}, \textit{U} maps. The polarization information catalog is created simultaneously from these Stokes maps following the procedures as described in Appendix~\ref{sec:pol}. Figure~\ref{fig:iqu} (b) and (c) show the final \textit{Q} and \textit{U}, (a) shows the final \textit{I} that is combined with \textit{Planck} 353 GHz flux via the J-comb algorithm \citep{Jiao2022} as described in Appendix~\ref{sec:j-comb}. 

\section{Results} \label{sec3}
\subsection{Dust Polarization Properties and Magnetic Field Morphology}
The projected plane-of-sky (POS) magnetic field orientations are derived by rotating the observed polarization pseudo-vectors by 90$^{\circ}$, based on the grain alignment assumption that the shortest axis of dust grains tends to align with the local magnetic field \citep{Lazarian2003}. The polarization pseudo-vectors are selected by criteria of $I/\delta_I \geq 10$, $PI/\delta_{PI} \geq 3$, and $\delta_p \leq 5\%$ where $\delta_I$ is the uncertainty of Stokes \textit{I}, $PI$ and $\delta_{PI}$ are the debiased polarized intensity and the corresponding uncertainty, $\delta_p$ is the uncertainty of polarization fraction. The inferred magnetic field orientations are shown in Figure~\ref{fig:pol_i} (a) overlaid on the column density ($N(\rm H_2)$) map, which is generated from level 2.5 processed archival \textit{Herschel} images by the J-comb algorithm as described in Appendix~\ref{sec:j-comb} \citep{Jiao2022}. The magnetic field is roughly perpendicular to the main filament structure with an average orientation of $60\pm20^{\circ}$\footnote{All position angles shown in this paper follow the IAU-recommended convention of measuring angles from the north towards the east.} (Figure~\ref{fig:pol_i} (c)).

Figure~\ref{fig:pol_i} (d) shows a decreasing polarization fraction trend with increasing initial dust emission intensity fitted with a power law index of $-0.73\pm0.04$. Figure~\ref{fig:pol_i} (e) exhibits the distribution of the polarization fraction, which peaks at $\sim$6.5\% with a tail extending to $\sim$15\%--25\%. The average and median of polarization fractions are $7.8\pm3.6$\%, and 7.0\%.%

As shown in Figure~\ref{fig:pol_i} (a), the magnetic field is roughly perpendicular to the filament structure. In general, the small-scale (14.4$\arcsec$, $\sim$0.13 pc) magnetic field traced by POL-2 agrees with the large-scale (4.8$\arcmin$, $\sim$2.6~pc) magnetic field traced by \textit{Planck}, showing similar average orientations, $60\pm20^{\circ}$ (POL-2) and $43\pm5^{\circ}$ (\textit{Planck}). However, in the center of the filament, the small-scale magnetic field shows a $\sim45^{\circ}$ difference from the large-scale one, suggesting the magnetic field orientation varies with increasing $N(\rm H_2)$, which may reflect the effects from gravity and turbulence (see Section~\ref{sec4} for further discussions).

\subsection{The magnetic field Strength}\label{sec:bstrength}
\begin{figure*}[ht!]
\centering
\includegraphics[scale=0.85]{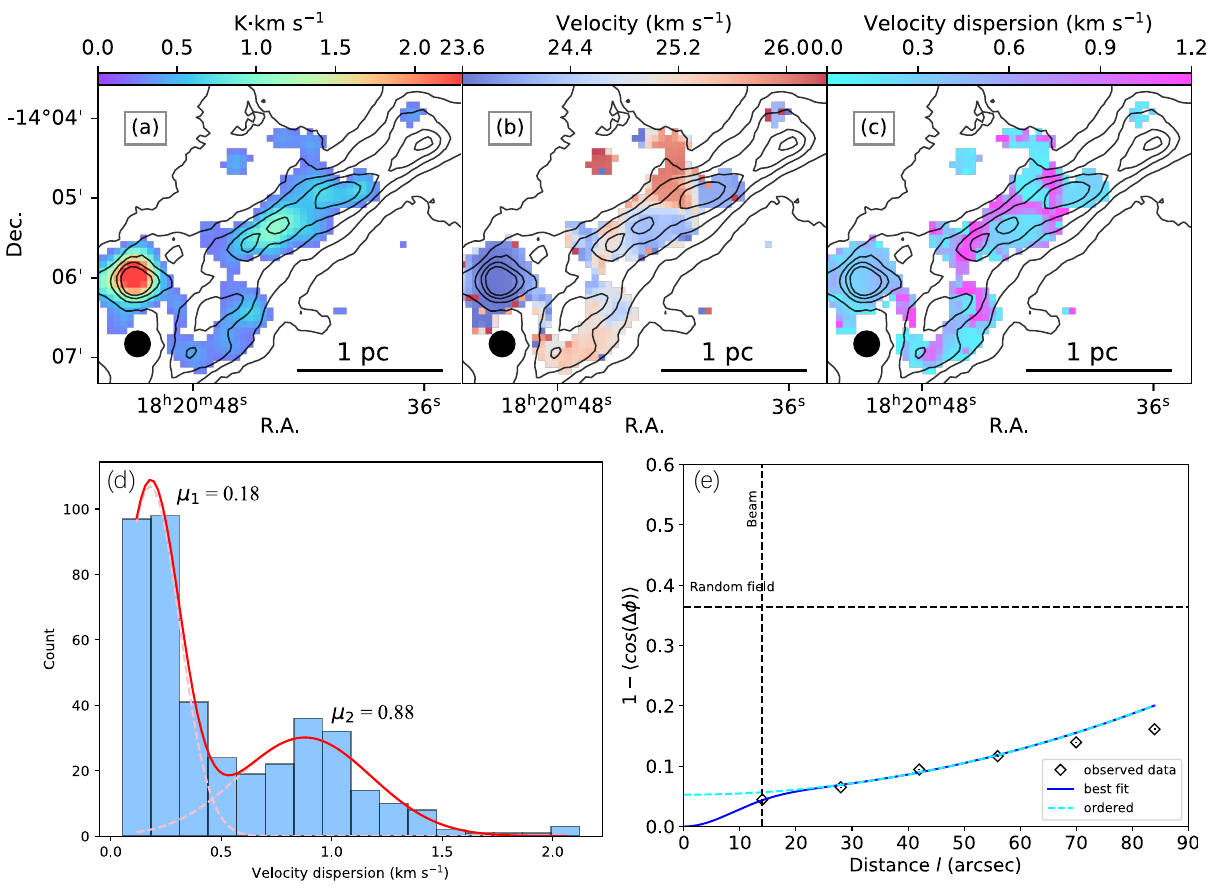}
\caption{\em{\textbf{(a)}: Integrated line emission of the isolated hyperfine component of $\rm N_2H^+$ with an S/N higher than 3. \textbf{(b)}: Centroid velocity map of $\rm N_2H^+$. \textbf{(c)}: Non-thermal velocity dispersion of $\rm N_2H^+$. Contours in (a-c) are the 850 $\mu$m Stokes I levels of [150, 200, 250, 300, 350]~mJy~beam$^{-1}$, the 18 $\arcsec$ beam size and a 1-pc scale bar are shown in the left and right lower corners, respectively. \textbf{(d)}: Distribution of the velocity dispersion shown in \textbf{(c)}. The red curve shows the best fitting of the bimodal distribution with peaks of $\sim0.18\,\rm km\, s^{-1}$ and $\sim0.88\,\rm km\, s^{-1}$, dashed curves represent the two single gaussian fittings. \textbf{(e)}: ADFs of G16.96+0.27. The diamond symbols represent the observed data points. Blue and cyan lines indicate the best-fitted result and the large-scale component of the best fit, respectively. The horizontal line marks the ADF value of a random field \citep[0.36,][]{LiuJH2021}.}}
\label{fig:mom_adf}
\end{figure*}

Before calculating the magnetic field strength, we estimate the gas density ($\rho$) and line-of-sight (LOS) non-thermal turbulent velocity dispersion ($\sigma_v$). As shown in Figrue~\ref{fig:pol_i} (a-b), we choose three cross-sections to apply gaussian-fit of $N(\rm H_2)$ profiles, and the average half maximum $N(\rm H_2)$ is $(2.38\pm0.17)\times10^{22}\rm\,cm^{-2}$ with an average \rm FWHM $\sim0.46\pm0.06$~pc. And most of the magnetic field segments are inside the contour of $N(\rm H_2)\sim2.38\times10^{22}\rm\,cm^{-2}$. Thus, we estimate the mass and $\rho$ using the $N(\rm H_2)$ map by assuming the filament within the $2.38\times10^{22}\rm\,cm^{-2}$ contour as a cylinder with a length of $\sim2.70\pm0.20$~pc and a diameter of $\sim0.46\pm0.06$~pc. As shown in Figure~\ref{fig:pol_i} (a), there are several small red contours not conjunct with the main structure and lack magnetic field segments, so we do not count them in for further calculations. Also, the protostellar core MM6 has the highest $N(\rm H_2)$ and strong ${\rm N_2H}^+ (J=1-0)$ (Figure~\ref{fig:mom_adf} (a)) emission but lacks magnetic field segments, so we mask MM6 to avoid bias when estimating the mass and the velocity dispersion as well. The mass and density are then calculated as $M\sim868_{-118}^{+116}\,\rm M_{\odot}$ and $\rho\sim1.31\times10^{-19}$~$\rm g\,cm^{-3}$, respectively, and further the line mass $M_l\sim321\pm43\,\rm M_{\odot}$~pc$^{-1}$. The corresponding volume density ($n_{\rm H_2}$) is derived as $\sim2.80\times10^4\,\rm cm^{-3}$ from $\rho = \mu_{\rm H_2}m_{\rm H}n_{\rm H_2}$, where $\mu_{\rm H_2}\simeq2.8$ is the molecular weight per hydrogen molecule \citep{Kauffman2008}, $m_{\rm H}$ is the atomic mass of hydrogen. 

We fit the FWHM line width, $\Delta v$, by hyperfine structure line fitting based on Nobeyama 45-m ${\rm N_2H}^+ (J=1-0)$ data \citep[Figure~\ref{fig:mom_adf} (a-c), adopted from][]{Tatematsu2021} with a resolution of 18$\arcsec$. The average $\sigma_v$ is then derived as $\sim0.52\,\rm km\, s^{-1}$ from $\Delta v/\sqrt{8\ln(2)} = \sqrt{\sigma_{\rm th}^{\phantom{th}2}+\sigma_v^{\phantom{v}2}}$, where $\sigma_{\rm th}=\sqrt{\frac{k_{\rm B}T}{m_{\rm N_2H^+}}}$ is the thermal velocity dispersion of $\rm N_2H^+$, $k_{\rm B}$ is the Boltzmann constant, $m_{\rm N_2H^+}sim4.85\times10^{-26}\,\rm kg$ is the molecular mass of $\rm N_2H^+$, $T$ is the dust temperature (the average value is $\sim$17.4~K with a standard deviation of $\sim$0.6~K) derived when generating the $N(\rm H_2)$ map by using the J-comb algorithm \citep{Jiao2022}. It is worth noting that as shown in Figure~\ref{fig:mom_adf} (d), $\sigma_v$ shows a bimodal distribution with peaks of $\sim0.18\,\rm km\, s^{-1}$ and $\sim0.88\,\rm km\, s^{-1}$, 
and the larger ones appear in the transition areas between the two velocity peaks shown in Figure~\ref{fig:mom_adf} (b), which may be a signature of two velocity components. If so, $\sigma_v$ is likely to be overestimated by a factor of 2 to 3. However, considering the data quality is insufficient for a deeper analysis, we still apply the average value $\sim0.52\,\rm km\, s^{-1}$ as $\sigma_v$. This may result in an overestimation of the strength of the magnetic field in the following analyses.

We apply the Davis–Chandrasekhar–Fermi (DCF) method \citep{Davis1951, CF1953a} to estimate the magnetic field strength. The DCF method relies on the following assumptions: the turbulence is isotropic; there is equipartition between the transverse turbulent magnetic field energy and kinetic energy; and the turbulent-to-ordered ($B_{\rm t}/B_{\rm o}$) or turbulent-to-total ($B_{\rm t}/B_{\rm tot}$) magnetic field ratio can be traced by the statistics of the magnetic field orientations. Then the ordered and total POS magnetic field strength could be estimated from
\begin{equation}
    B_{\rm o} = f_{\rm dcf}\sqrt{4\pi\rho}\frac{\sigma_v}{B_{\rm t}/B_{\rm o}},
\end{equation}
and
\begin{equation}
    B_{\rm tot} = f_{\rm dcf}\sqrt{4\pi\rho}\frac{\sigma_v}{B_{\rm t}/B_{\rm tot}},
\end{equation}
where $f_{\rm dcf}$ is the correction factor. When the ordered magnetic field is prominent, $B_{\rm t}/B_{\rm o}$ and $B_{\rm t}/B_{\rm tot}$ are usually estimated from $B_{\rm t}/B_{\rm o} \sim B_{\rm t}/B_{\rm tot} \sim \sigma_{\theta}$, where $\sigma_{\theta}$ is the angular dispersion of POS magnetic field orientations.

We note that there are many versions of the DCF method showing different ways to quantify $B_{\rm t}/B_{\rm tot}$ more accurately \citep[e.g.][]{Falceta2008, CY2016, LiuJH2021}. Here we use two of them to estimate the magnetic field strength for comparison and analysis: the classical DCF method \citep{Ostriker2001}, and the calibrated angular dispersion function (ADF) method \citep{Hildebrand2009, Houde2009, Houde2016}, a modified DCF method. Using the $f_{\rm dcf} = 0.5$ derived from the numerical models \citep{Ostriker2001} and the estimated $\sigma_{\theta}$ of $20\pm1^{\circ}$, we obtain $B_{\rm dcf}\sim0.5\sqrt{4\pi\rho}\sigma_v/\sigma_{\theta}\sim96\pm17\,\rm\mu G$. 

\cite{LiuJH2021} calibrated the ADF method and found it accounts for the ordered magnetic field structure and beam smoothing. The turbulent correlation effect is derived from 
\begin{equation}\label{eq:adf}
1-\langle\cos[\Delta \Phi(l)]\rangle \simeq\frac{\langle B_t^{\phantom{t}2}\rangle}{\langle B^2\rangle}\times(1-e^{-l^2/2(l_{\delta}^2+2W^2)})+a'_2l^2,
\end{equation}
where $\Delta \Phi(l)$ is the angular difference of two magnetic field angles separated by a distance of $l$, $l_{\delta}$ is the turbulent correlation length for local turbulent magnetic field, $W=l_{\rm beam}/\sqrt{8\ln2}$ is the standard deviation of the Gaussian beam size and the second-order term $a'_2l^2$ is the first term of Taylor expansion of the ordered component of ADF. Figure~\ref{fig:mom_adf} (e) shows the ADF of G16.96+0.27, we fit ADF by reduced $\chi^2$ minimization with the best-fitted $(\langle B_t^{\phantom{t}2}\rangle/\langle B^2\rangle)^{0.5}$ of 0.23, thus by using $B\sim0.21\sqrt{4\pi\rho}\sigma_v (\langle B_t^{\phantom{t}2}\rangle/\langle B^2\rangle)^{-0.5}$ \citep{LiuJH2021}, we obtain $B_{\rm adf}\sim60\pm10\,\rm\mu G$. Thus, we estimate an average strength of $B_{\rm pos}=0.5(B_{\rm dcf}+B_{\rm adf})\sim78\pm20\,\rm\mu G$ for further analysis.

\begin{figure}[ht!]
\centering
\includegraphics[scale=0.6]{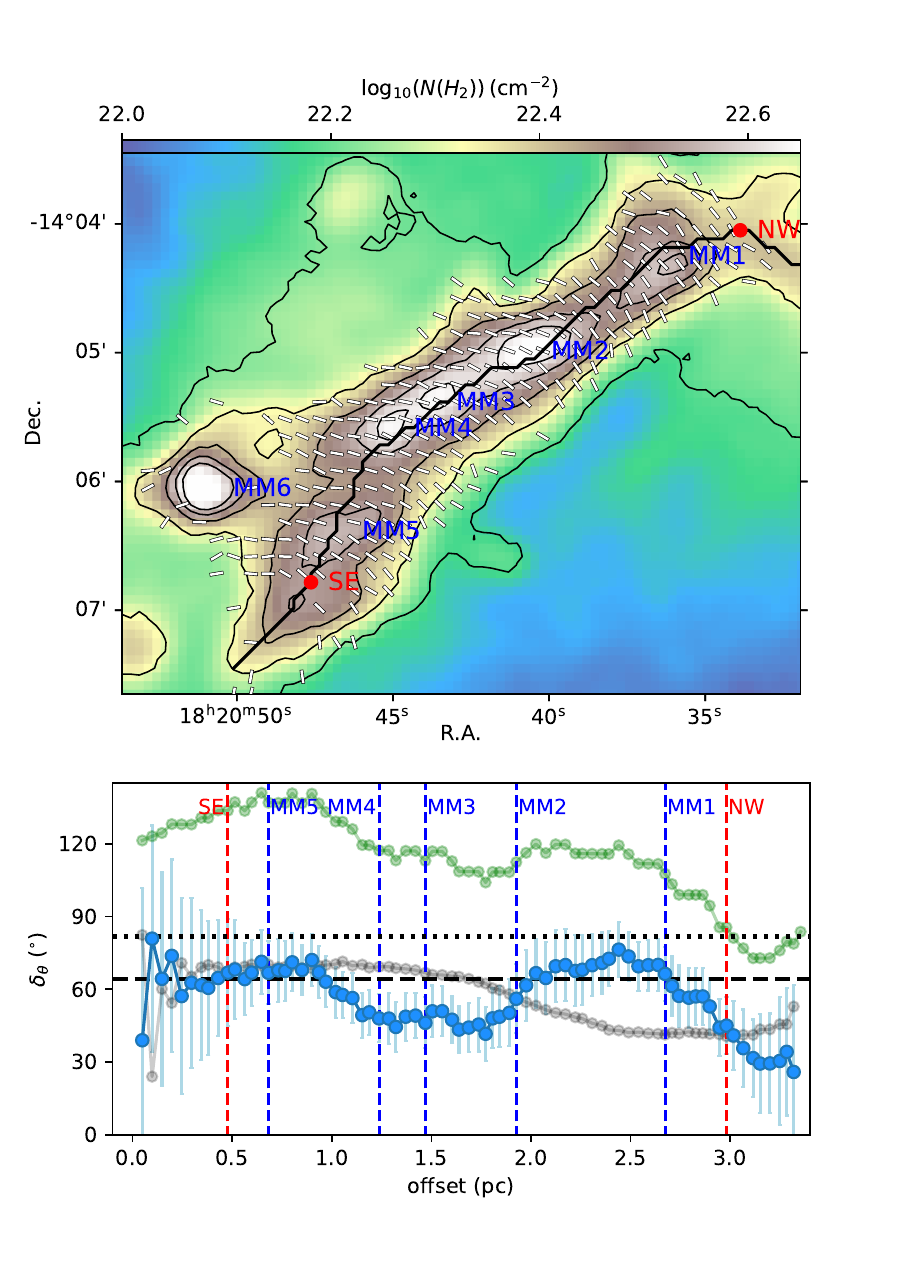}
\caption{\em{\textbf{Upper}: magnetic field orientation map overlaying the $N(\rm H_2)$ map with contours showing 850~$\mu$m Stokes I levels of [150, 200, 250, 300, 350]~mJy~beam$^{-1}$. The black curve shows the filament skeleton derived from \textit{FilFinder}. The two red points mark the two ends of the skeleton. The four fragments are marked as MM1-6. \textbf{Lower}: The filament skeleton and magnetic field angle difference. The blue curve shows the angle difference between the local magnetic field and the filament skeleton, the offset is counted from the southeast to the northwest of the skeleton. The grey and green curves are the magnetic field orientation and filament orientation along the skeleton, respectively. The dashed black line marks the angle difference between the mean filament skeleton orientation and the mean POL-2 magnetic field orientation, 65.7$^{\circ}$. In contrast, the dotted black one shows the angle difference between the mean filament skeleton orientation and the mean \textit{Planck} magnetic field orientation, 81.8$^{\circ}$. The two red dashed lines mark the locations of the two endpoints in the upper panel and the four blue dashed lines mark the locations of MM1-5.}}
\label{fig:fil_b}
\end{figure}

\begin{figure*}[ht!]
\centering
\includegraphics[scale=0.475]{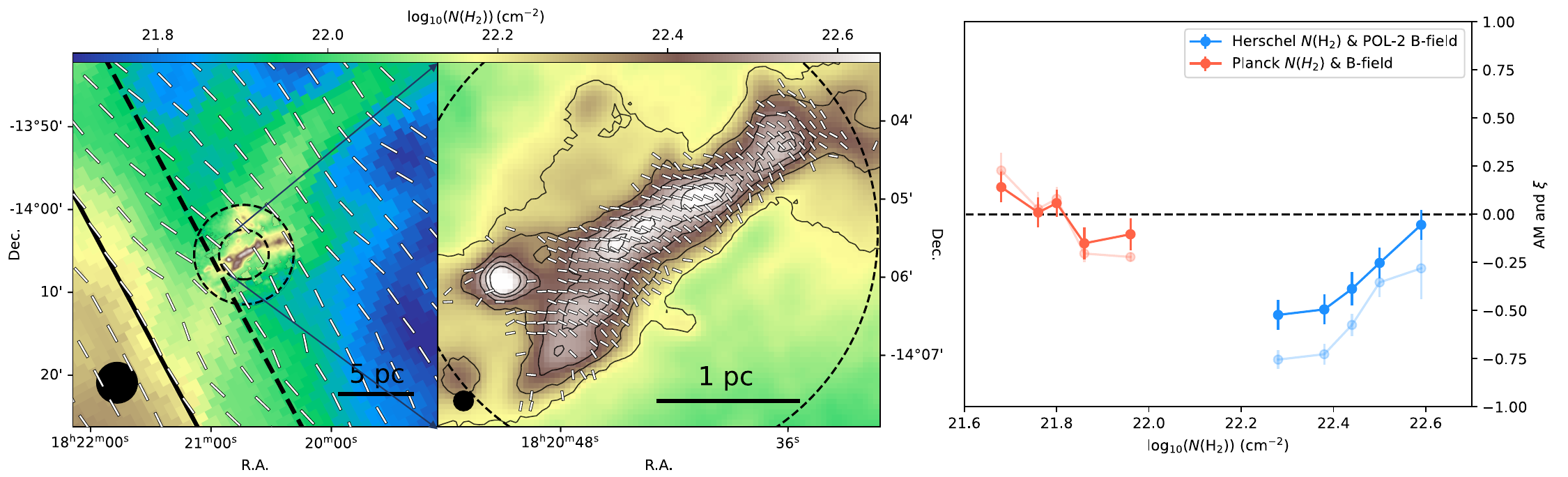}
\caption{\em{\textbf{Left}: column density inferred from \textit{Planck} dust optical depth map at 353GHz ($\tau_{353}$) overlaid with the one (within the outer black dashed circle) inferred from \textit{Herschel} data by using the J-comb algorithm\citep{Jiao2022}. The segments show the magnetic fields derived from \textit{Planck} 353 GHz polarization data, the black circle shows the beam of \textit{Planck} 353GHz data with a resolution of 4.8$\arcmin$, the black solid line and dashed line represent 0$^{\circ}$ and 0.185$^{\circ}$ galactic latitude, respectively. The two dashed circles mark the central regions of 12$\arcmin$ and 6$\arcmin$ radius, respectively. A 5-pc scale bar is shown in the lower right corner. \textbf{Middle}: column density inferred from \textit{Herschel} data by using the J-comb algorithm with a resolution of 18$\arcsec$ (the beam is shown in the lower left corner), overlaid segments represent the magnetic field orientation inferred from JCMT/POL-2 850~$\mu$m polarization data. The dashed circle marks the central 6$\arcmin$ region, which has a useful level of coverage in the POL-2 observations. The contours show the 850~$\mu$m Stokes I levels of [150, 200, 250, 300, 350]~mJy~beam$^{-1}$. A 1-pc scale bar is shown in the lower right corner. \textbf{Right}: Alignment measure parameter $AM$ and HRO parameter $\xi$ calculated for the different $N(\rm H_2)$ bins in G16.96+0.27. Red data points show the $AM$ values (light red for $\xi$) inferred from \textit{Planck} 353 GHz data, and blue ones show the $AM$ values (light blue for $\xi$) inferred from \textit{Herschel} $N(\rm H_2)$ and POL-2 magnetic field.}}
\label{fig:cd_am}
\end{figure*}

\section{Discussion} \label{sec4}
As G16.96+0.27 has a relatively simple filamentary shape, for further analysis, we identify the skeleton of this structure by applying the FilFinder algorithm \citep{2015Filfinder} to the $N(\rm H_2)$ map. As shown in the upper panel of Figure~\ref{fig:fil_b}, we mask MM6 when finding the skeleton for the reasons mentioned in Section~\ref{sec3}.

\subsection{Equilibrium State}\label{sec:4.1}
For an unmagnetized filamentary cloud, the virial mass per unit length is $M_{{\rm vir},l} = 2\sigma_{\rm tot}^{\phantom{tot}2}/G$ \citep{Fiege2000}, where $\sigma_{\rm tot} = \sqrt{c_{\rm s}^{\phantom{s}2}+\sigma_v^{\phantom{v}2}}$ is the total velocity dispersion, $c_{\rm s}=\sqrt{\frac{k_{\rm B}T}{\mu_p m_{\rm H}}}\simeq0.25\, \rm km\, s^{-1}$ is the isothermal sound speed with an average \textit{T} of $\sim17.4\rm\, K$ and $\mu_p\simeq2.37$ is the mean molecular weight per free particle \citep{Kauffman2008}. The virial parameter is then defined as 
\begin{equation}
    \alpha_{\rm vir} = \frac{M_{{\rm vir},l}}{M_l} = \frac{2\sigma_{\rm tot}^{\phantom{tot}2}}{G M_l},
\end{equation}
$M_{{\rm vir},l}$ of G16.96+0.27 is $\sim153\,\rm M_{\odot}pc^{-1}$, and thus $\alpha_{\rm vir}\sim0.48\pm0.07$, suggesting that turbulence is weaker than gravity. 

Taking the magnetic field into account, the maximum mass per length that the magnetic field can support against gravity is $M_{\Phi,l} = \Phi_l/(2\pi\sqrt{G})$, where $\Phi_l$ is the magnetic flux per unit length. The local magnetic stability critical parameter \citep{Crutcher2004} is then defined as 
\begin{equation}
    \lambda = \frac{M_l}{M_{\Phi,l}} = \frac{\mu_{\rm H_2}m_{\rm H}N({\rm H_2})}{B/(2\pi\sqrt{G})}\simeq7.6\times10^{-21}\frac{N({\rm H_2})}{B},
\end{equation}
where $N({\rm H_2})$ is the column density in units of $\rm cm^{-2}$ and \textit{B} is the total 3D magnetic field strength in units of $\rm\mu G$. For G16.96+0.27, the average $ N(\rm H_2)$ is $\sim(3.10\pm0.52)\times10^{22}\rm cm^{-2}$, $B_{\rm pos}\sim78$~$\rm\mu G$, considering $B = \frac{4}{\pi}\overline{B_{\rm pos}}\sim 99~\rm\mu G$ \citep{Crutcher2004}, $\lambda$ is derived as $\sim2.56\pm0.74$, indicating the magnetic field is also weaker than gravity. However, \cite{Crutcher2004} proposed that the observed $M/\Phi_l$ will be overestimated by up to a factor of 3 due to geometrical effects, this correction results in a lower limit of $\lambda$ as $\lambda_{\rm min}\sim0.86\pm0.25$, showing a possibility of the magnetic field being stronger than gravity at some certain inclination angles.

\cite{Kashiwagi2021} found that when a filamentary cloud is supported by both a perpendicular magnetic field and thermal and turbulent motions, the maximum stable mass per unit length is $M_{{\rm crit},l}\simeq\sqrt{M_{\Phi,l}^{\phantom{\Phi,l}2}+M_{{\rm vir},l}^{\phantom{vir,l}2}}$, which implies
\begin{equation}
    \frac{M_l}{M_{{\rm crit},l}} = \frac{1}{\sqrt{\lambda^{-2}+\alpha_{vir}^2}}.
\end{equation}
$(M_l/M_{{\rm crit},l})$ is $\sim1.62\pm0.23$, suggesting magnetic field and turbulence together are weaker than gravity. However the lower limit $(M_l/M_{{\rm crit},l})_{\rm min}\sim0.79\pm0.20$ if applying $\lambda_{\rm min}\sim0.86$, suggesting magnetic field and turbulence together are stronger than gravity. Thus, a $(M_l/M_{{\rm crit},l})\sim1$ is more convincing, indicating G16.96+0.27 is in a quasi-equilibrium state. However, considering the magnetic field strength could be overestimated (see Section~\ref{sec:bstrength}), $(M_l/M_{{\rm crit},l})$ could be even larger and thus the filament is more likely to be a gravitationally bound system. 

\subsection{Fragmentation inside the Quiescent Filament}\label{sec:spacing}
As shown in Figure~\ref{fig:iqu} (a-c), there are several possible fragments along the filament major axis, which have been identified as protostellar (MM1, MM2 and MM6) and starless (MM3, MM4 and MM5) cores \citep{Kim2020, Tatematsu2021, Mannfors2021}. Such fragmentation can be explained by the so-called 'sausage instability' of a cylindrical gas structure \citep[e.g.][]{CF1953b, Inutsuka1992, Wang2014, Contreras2016}. In an isothermal gas cylinder with a helicoidal magnetic field, \cite{Nakamura1993} predicted a typical spacing of fragments by
\begin{equation}\label{eq:spacing}
   L\simeq\frac{2\pi}{0.72}H[(1+\gamma)^{1/3}-0.6]^{-1},
\end{equation}
where $\gamma = B_{\rm c}^{\phantom{\rm c}2}/(8\pi\rho_{\rm c}\sigma^2)$, $\rho_{\rm c}$ and $B_{\rm c}$ are the density and magnetic field strength in the center, \textit{H} is the scale height. According to \cite{Nakamura1993}, for a cylindrical gas structure with a magnetic field of $\textbf{B} = (0, B_{\phi}, B_{\rm z})$, the scale height (\textit{H}) is defined from
\begin{equation}
    4\pi G\rho_{\rm c}H^2 = \sigma^2+\frac{B_{\rm c}^{\phantom{\rm c}2}}{16\pi\rho_{\rm c}}(1+\cos^2{\theta}),
\end{equation}
where $\theta = \lim\limits_{r\to\infty}\tan^{-1}{B_{\phi}/B_{\rm z}}$ denotes the ratio of $B_{\phi}$ and $B_{\rm z}$, when $\theta = 0$, the magnetic field is parallel to the filament. $\sigma = c_{\rm s}\sim0.25\, \rm km\, s^{-1}$ is the isothermal sound speed, and it is replaced by $\sigma_{\rm v}\sim0.52\, \rm km\, s^{-1}$ when the fragmentation is governed by the turbulence rather than thermal Jeans instability. In G16.96+0.27, we estimated $\theta$ as the angle between the mean POS magnetic field orientation and filament skeleton ($\sim65.7^{\circ}$), and $\rho_{\rm c}\sim1.92\times10^{-19}$~$\rm g\,cm^{-3}$ from $\rho_{\rm c} = \frac{\rho n_{\rm H_2,c}}{n_{\rm H_2}}$, where $n_{\rm H_2,c}\sim4.55\times10^{22}\rm\,cm^{-2}$ is the column density in the center. And we assume $B_{\rm c}\sim B\sim99\,\rm\mu G$. Therefore, we have thermal support $H_{\rm thermal}\sim0.04$~pc and turbulent support $H_{\rm turbulent}\sim0.05$~pc. Further, $L_{\rm thermal}$ of $\sim0.31$~pc and $L_{\rm turbulent}$ of $\sim0.75$~pc.

The separations between two nearby fragments are counted from center to center with an average of $\sim0.47\pm0.15$~pc (range in $\sim0.21-0.65$~pc). Considering the possible effect of projection, the separations would be $2/\pi$ times of the 3D ones on average \citep{Sanhueza2019}. And the possible 3D separations are $\sim0.33-1.02$~pc with an average of $\sim0.74\pm0.23$~pc, which favors turbulent support rather than thermal support. It is worth noting that the popular-used spacing equation $L\simeq22H$ with $H=\sigma\sqrt{4\pi G\rho_{\rm c}}$ is under the condition of $\gamma = 0$ (i.e. ignorance of the magnetic field). Though Equation~\ref{eq:spacing} cares about the magnetic field, it is under a helicoidal magnetic field assumption. For G16.96+0.27, we have no evidence of a helicoidal magnetic field, thus the result of \textit{L} is for reference only.

\subsection{Multiscale Cloud-field Alignment}\label{sec:c-f alignment}
As mentioned in Section~\ref{sec3}, the $\sim0.13$ pc scale magnetic field shows rough agreement with the $\sim2.6$ pc-scale one but with some discrepancies in high-density regions. For further analysis, we use the histogram of relative orientations \citep[HROs,][]{Soler2013, Soler2017} method. The HRO parameter $\xi$ is defined as
\begin{equation}\label{eq:hro}
    \xi = \frac{A_0-A_{90}}{A_0+A_{90}},
\end{equation}
where $A_0$ and $A_{90}$ represent the areas under the histogram of $\Phi$ (the angle between the POS magnetic field orientation and the iso-column density structure) value from $0^{\circ}$ to $22.5^{\circ}$ and $67.5^{\circ}$ to $90^{\circ}$, respectively. While the derivation of $\xi$ completely ignores angles from $22.5^{\circ}$ to $67.5^{\circ}$, \cite{Jow2018} improved the HRO analysis with projected Rayleigh statistic (PRS) to overcome the shortcoming. Thus, we also use the normalized version of PRS, the alignment measure (AM) parameter \citep{Lazarian2018, LiuJH2022} to study the relative alignment. The AM parameter is defined as
\begin{equation}\label{eq:am}
    {\rm AM} = <\cos{2\delta_{\theta}}>,
\end{equation}
where $\delta_{\theta}$ represents the relative orientation angle between the magnetic field and gas structure. A positive value of AM and $\xi$ means the magnetic field tends to be parallel with $N(\rm H_2)$ contour while a negative value stands for a perpendicular alignment.


We use POL-2 magnetic field (14.4\arcsec) data and $N(\rm H_2)$ from the J-comb algorithm (18\arcsec) to calculate the small-scale AM and $\xi$. For the large-scale ones, we derive the column density by $\tau_{353}/N(\rm H) = 1.2\times10^{-26}\rm cm^{-2}$ to match the \textit{Planck} 353 GHz magnetic field data, where $\tau_{353}$ is \textit{Planck} 353 GHz dust optical depth \citep{PlanckXI}. We mask regions with galactic latitude lower than 0.185$^{\circ}$ to avoid the effect of the emission from the Galactic plane. 

As shown in the right panel of Figure~\ref{fig:cd_am}, AM and $\xi$ exhibit the same behavior with the increasing $N(\rm H_2)$. They go from positive to negative at large-scale (red curve, low $N(\rm H_2)$ traced by \textit{Planck} data) but have the opposite behavior at small-scale (blue curve, high $N(\rm H_2)$ traced by \textit{Herschel}), showing a positive slope. 

The behavior of AM and $\xi$ indicates the following cloud-field alignment phenomenon. In the diffuse environment (with a $N(\rm H_2)$ of $\sim5.0\times10^{21}\rm cm^{-2}$, shown in light-blue in the left panel of Figure~\ref{fig:cd_am}), the magnetic field tends to be parallel with the gas structure. In the host structure (with a $N(\rm H_2)$ of $\sim1.3\times10^{22}\rm cm^{-2}$, shown in green in the left panel of Figure~\ref{fig:cd_am}), the magnetic field turns to be perpendicular to the gas structure, which is perpendicular to the galactic plane and extends to the northwest conjunct with M16, an active high-mass star-forming cloud \citep[e.g.][]{Hester1996, Sugitani2002, Pattle2018}. In the outskirts (with a $N(\rm H_2)$ of $\sim2.5\times10^{22}$~$\rm cm^{-2}$) of the main structure (the middle panel of Figure~\ref{fig:cd_am}), the alignment keeps perpendicular as in the host structure. In the dense center (with a $N(\rm H_2)$ higher than $\sim3.2\times10^{22}$~$\rm cm^{-2}$), it goes to be $\sim45^{\circ}$ and shows a possible trend to be closer to parallelism. 

Along the filament, we average the magnetic field orientation and skeleton orientation using a 32\arcsec\, filter, as 32\arcsec($\sim0.29$ pc) is similar to the diameter of the filament, and calculate $\delta_\theta$ between them. As shown in the lower panel of Figure~\ref{fig:fil_b}, regions beyond points SE and NW have few magnetic field segments and substantial uncertainties on $\delta_\theta$, so we mainly discuss the skeleton in-between. $\delta_\theta$ shows fluctuations along the skeleton, it is $\sim65$--$80^{\circ}$ at both ends, but becomes $\sim40$--$50^{\circ}$ at the center. The behavior of $\delta_\theta$ indicates that the local magnetic field tends to be perpendicular to the filament at the two diffuse ends but shows a trend to have a smaller offset with the filament at the dense center, which agrees with the behavior of AM and $\xi$ shown in Figure~\ref{fig:cd_am}. 

\cite{Pillai2020} found a similar positive slope of $\xi$ in dense regions of the hub-filament system Serpens South, suggesting the gas filaments merging into the central hub and reorienting the magnetic field in the dense gas flows.  \cite{Kwon2022} found that $\xi$ of Serpens Main shows large fluctuations with increasing $N(\rm H_2)$, which is interpreted as the density gradient along the elongated structures becoming significant and magnetic field being dragged along with the increasing density. Furthermore, in the massive IRDC G28.34 (the Dragon cloud), \cite{LiuJH2024} found that the AM parameter also goes from negative to positive with increasing $N(\rm H_2)$, and fluctuates around 0 in the very dense region traced by ALMA, suggesting G28.34 is located in a trans-to-sub-Alfv\'{e}nic environment. 

However, compared to Serpens South, Serpens Main and G28.34, G16.96+0.27 has a much simpler structure and less star-formation activity. As shown in the upper panel of Figure~\ref{fig:iqu}, MM1 and MM6 are not connected with the main structure from the infrared view, and most of the rest (MM3-MM5) are still starless cores. All the fragments have an average POS separation of $\sim0.47$~pc, which may favor turbulent-supported fragmentation rather than thermal-supported one (see the discussion in Section~\ref{sec:spacing}). These results may indicate the following phenomenon: The G16.96+0.27 filament as a whole is quiescent and in quasi-equilibrium as reflected by the dark infrared morphology shown in Figure~\ref{fig:iqu} and $M_l/M_{{\rm crit},l}$ value. However, in the center region, the star formation process may have already begun as reflected by the fragments shown in Figure~\ref{fig:iqu} and discussion in Section~\ref{sec:spacing}. Thus gravity has overcome the support of the magnetic field and turbulence, and dragged the field lines to align with the filament structure as has been observed at smaller scales in different targets \citep[e.g.,][]{Sanhueza2021, Cortes2024}. Further, higher-resolution observations toward the center region are needed to investigate whether AM and $\xi$ would fluctuate around 0 like G28.34 \citep{LiuJH2024} or turn around showing the decreasing trend seen in the very dense regions ($N(\rm H_2)\geq1.6\times10^{23}$~$\rm cm^{-2}$) in Serpens Main \citep{Kwon2022}. 

\section{Summary} \label{sec5}
In this paper, we have presented the JCMT/POL-2 polarization observations towards an IRDC, G16.96+0.27, and the main conclusions are as follows:

(1) The average magnetic field orientation traced by JCMT/POL-2 is $\sim60^{\circ}$, and a significant number of magnetic field segments exhibit a perpendicular alignment with the filament structure of G16.96+0.27, with an average angle difference of $\sim66^{\circ}$. This result is consistent with the larger-scale magnetic field orientations traced by \textit{Planck} 353~GHz data. The POS magnetic field strength is estimated to be $B_{\rm pos,dcf}\sim96$~$\rm\mu G$ and $B_{\rm pos,adf}\sim60$~$\rm\mu G$ using the classical DCF method and the ADF method, with an average strength of $\sim78\rm\mu G$. 

(2) The virial parameter and magnetic stability critical parameter are calculated as $\alpha_{\rm vir}\sim0.48$, $\lambda\sim2.56$ with $\lambda_{\rm min}\sim0.86$, respectively. And the estimated $(M_l/M_{{\rm critc},l})$ is $\sim1$, indicating that G16.96+0.27 is in a quasi-equilibrium state, but is more likely to be a gravitationally bound status when considering the magnetic field could be overestimated.

(3) We calculate the HRO parameter $\xi$ and AM parameter based on \textit{Planck} and JCMT data to study multiscale cloud-field alignment. With increasing $N(\rm H_2)$, they first go across 0 to a negative minimum and then move back to 0. Along the filament, we apply the FilFinder algorithm to identify the skeleton of G16.96+0.27 and find that the local cloud-field alignment varies along the filament. The alignment is perpendicular at both diffuse ends but turns to be $\sim45^{\circ}$ at the dense center, which is consistent with the behavior of AM and $\xi$. We also find that the observed separations among the fragments are in agreement with the predicted spacing from the 'sausage' instability theory under turbulent support assumption (Appendix~\ref{sec:spacing}). These results may reflect that although G16.96+0.27 is in quasi-equilibrium overall, fragmentation has already begun in the center of the filament, and such a phenomenon of cloud-field alignment inside IRDCs may be a possible sign of an early stage of star formation activity.

\section*{Acknowledgements}
This work has been supported by the National Key R\&D Program of China (No.\ 2022YFA1603100), Shanghai Rising-Star Program (23YF1455600), and Natural Science Foundation of Shanghai (No.\ 23ZR1482100). T.L.\ acknowledges support from the National Natural Science Foundation of China (NSFC), through grants No.\ 12073061 and No.\ 12122307, the Tianchi Talent Program of Xinjiang Uygur Autonomous Region, and the international partnership program of the Chinese Academy of Sciences, through grant No.\ 114231KYSB20200009. X.L.\ acknowledges support from the NSFC through grant Nos.\ 12273090 and 12322305, and the Chinese Academy of Sciences (CAS) ``Light of West China'' Program No.\ xbzg-zdsys-202212. MJ acknowledges the support of the Research Council of Finland Grant No. 348342. C.W.L. acknowledges support from the Basic Science Research Program through the NRF funded by the Ministry of Education, Science and Technology (NRF-2019R1A2C1010851) and by the Korea Astronomy and Space Science Institute grant funded by the Korea government (MSIT; project No. 2024-1-841-00). W.K. was supported by the National Research Foundation of Korea (NRF) grant funded by the Korea government (MSIT; project No. RS-2024-00342488). K.P. is a Royal Society University Research Fellow, supported by grant number URF\textbackslash R1\textbackslash 211322. PS was partially supported by a Grant-in-Aid for Scientific Research (KAKENHI Number JP22H01271 and JP23H01221) of JSPS.

The James Clerk Maxwell Telescope is operated by the East Asian Observatory on behalf of The National Astronomical Observatory of Japan; Academia Sinica Institute of Astronomy and Astrophysics; the Korea Astronomy and Space Science Institute; the National Astronomical Research Institute of Thailand; Center for Astronomical Mega-Science (as well as the National Key R\&D Program of China with No.\ 2017YFA0402700). Additional funding support is provided by the Science and Technology Facilities Council of the United Kingdom and participating universities and organizations in the United Kingdom and Canada. Additional funds for the construction of SCUBA-2 were provided by the Canada Foundation for Innovation.


%



\software{Astropy \citep{2013Astropy,2018Astropy}, FilFinder \citep{2015Filfinder}
          }



\appendix
\section{Data reduction procedures of JCMT/POL-2 data}\label{sec:pol}
As the polarization fraction is forced to be positive, a bias is thus introduced \citep{Vaillancourt2006}, and the therefore debiased polarized intensity (\textit{PI}) and corresponding uncertainty ($\delta_{PI}$) are calculated from
\begin{equation}
    PI = \sqrt{Q^2+U^2-0.5(\delta_Q^{\phantom{Q}2}+\delta_U^{\phantom{U}2})},
\end{equation}
and
\begin{equation}
    \delta_{PI} = \sqrt{\frac{Q^2\delta_Q^{\phantom{Q}2}+U^2\delta_U^{\phantom{U}2}}{Q^2+U^2}},
\end{equation}
where $\delta_Q$ and $\delta_U$ are the uncertainties of \textit{Q} and \textit{U}. The debiased polarization fraction (\textit{p}) and corresponding uncertainty ($\delta_p$) are then derived by
\begin{equation}
    p = PI/I,
\end{equation}
and
\begin{equation}
    \delta_p = \sqrt{\frac{\delta_{PI}^{\phantom{PI}2}}{I^2}+\frac{PI^2\delta_I^{\phantom{I}2}}{I^4}},
\end{equation}
where $\delta_I$ is the uncertainty of \textit{I}. Next, polarization angle ($\theta$) and corresponding uncertainty ($\delta_{\theta}$) are calculated \citep{Naghizadeh-Khouei1993} from
\begin{equation}
    \theta = 0.5\tan^{-1}(U/Q),
\end{equation}
and
\begin{equation}
    \delta_{\theta} = \frac{1}{2}\sqrt{\frac{Q^2\delta_U^{\phantom{U}2}+U^2\delta_Q^{\phantom{Q}2}}{(Q^2+U^2)^2}}.
\end{equation}

\section{J-comb algorithm}\label{sec:j-comb}
We get the column density map by applying the J-comb algorithm \citep{Jiao2022} based on level 2.5 processed, archival \textit{Herschel} data and this JCMT 850~$\mu$m data, the main procedures are as follows:

1. Derive combined Stokes \textit{I} map. We extrapolated an 850~$\mu$m flux map from the spectral energy distribution (SED) of \textit{Herschel} 250/350/500~$\mu$m data using the Spectral and Photometric Imaging REceiver \citep[SPIRE; obsID: 1342228342;][]{Griffin2010}. Taking this map as a model image, we deconvolved the \textit{Planck} 353 GHz map \citep{PlanckXIX} with the Lucy-Richardson algorithm \citep{Lucy1974}. The obtained deconvolved map has an angular resolution close to the SPIRE 500 $\mu$m data and preserves the flux level of the initial \textit{Planck} map. Then the 
combined Stokes I map (as shown in Figure~\ref{fig:iqu} (a)) was generated via the J-comb algorithm \citep{Jiao2022} by combining the deconvolved map with JCMT 850~$\mu$m Stokes \textit{I} map in the Fourier domain.

2. SED fitting. We smoothed \textit{Herschel} images at 70/160~$\mu$m using the Photodetector Array Camera and Spectrometer \citep[obsID: 1342228372;][]{Poglitsch2010}, SPIRE 250~$\mu$m and the combined JCMT 850~$\mu$m Stokes \textit{I} map to a common angular resolution of the largest beam. We weighted the data points by the measured noise level in the least-squares fits. As a modified blackbody assumption, the flux density $S_{\nu}$ at the frequency $\nu$ is given by
\begin{equation}
    S_{\nu} = \Omega_m B_{\nu}(T)(1-e^{-\tau_{\nu}}),
\end{equation}
where $\Omega_m$ is the solid angle, $B_{\nu}(T)$ is the Planck function at temperature $T_{\rm dust}$. Then the column density is derived from 
\begin{equation}
    N(\rm H_2) = \tau_{\nu}/\kappa_{\nu}\mu m_{\rm H},
\end{equation}
where $\kappa_{\nu}=0.1{\rm cm^2 g^{-1}}(\nu/1000 GHz)^{\beta}$ is the dust opacity assuming a gas-to-dust ratio of 100 and an opacity index $\beta$ of 2 \citep{Hildebrand1983, Beckwith1990}, $\mu=2.8$ is the molecular weight per hydrogen molecule \citep{Kauffman2008}, and $m_{\rm H}$ is the atomic mass of hydrogen.


\bibliography{sample631_arxiv.bbl}{}
\bibliographystyle{aasjournal}


\end{CJK}
\end{document}